\documentclass[aps,prd,12pt,nofootinbib]{revtex4-1}
\usepackage{amsmath,amssymb}

\def\zi{\mathbb{Z}}

\begin{document}
\title{Non-geometric Calabi-Yau compactifications\\ and fractional mirror symmetry}
\author{Dan \surname{Isra\"el}}
\email{israel@lpthe.jussieu.fr}
\affiliation{Sorbonne Universit\'es, UPMC Univ Paris 06, UMR 7589, LPTHE, F-75005, 
Paris, France}
\affiliation{CNRS, UMR 7589, LPTHE, F-75005, Paris, France}

\begin{abstract}
We construct a wide class of non-geometric compactifications of type II superstring theories preserving N=1 
space-time supersymmetry in four dimensions, starting from Calabi-Yau compactifications at Gepner points. Particular 
examples of this construction provide quantum equivalences between Calabi-Yau compactifications and non-Calabi-Yau ones, 
generalizing mirror symmetry. The associated Landau-Ginzburg models involve both chiral and twisted chiral 
multiplets hence cannot be lifted to ordinary Calabi-Yau gauged linear sigma-models. 
\end{abstract}
\maketitle

\section{Introduction}

It is widely acknowledged that Calabi-Yau (CY) manifolds form only a small subset of supersymmetric string compactifications. 
Understanding more general compactifications is an important goal, both for probing the 
quantum geometry of string theory and for obtaining four-dimensional models with fewer moduli and fewer supersymmetries. Besides 
compactifications with Ramond-Ramond fluxes, that are quite successful in this respect but lack a usable worldsheet formulation, 
it is desirable to find models with a better grip on $\alpha'$ corrections beyond the supergravity regime. Unlike in heterotic strings, 
it is not possible to consider type II compactifications with NSNS three-form flux only, because of the tadpole condition 
$\int e^{-2\Phi} H\wedge \star H=0$ coming from the equations of motion. 

It leaves the possibility of using non-geometric fluxes; compactifications of this type  
have been described as asymmetric orbifolds of rational tori~\cite{Narain:1986qm,Dabholkar:1998kv,Anastasopoulos:2009kj}, 
using free-fermion models~\cite{Kawai:1986ah,Antoniadis:1986rn} or as T-folds that are locally geometric and their (generalized) 
T-duals~\cite{Dabholkar:2002sy,Hull:2004in,Hellerman:2002ax}. Studying such 
non-geometric fluxes in {\it interacting} rather than free worldsheet conformal field theories (CFTs) would allow to 
understand how non-geometric compactifications can be defined in non-trivial backgrounds. 
A large class of supersymmetric compactifications on Calabi-Yau manifolds, in the stringy regime of negative K\"ahler moduli, are 
described by superconformal field theories constructed by Gepner using $\mathcal{N}=(2,2)$ 
minimal models as building blocks~\cite{Gepner:1987qi,Gepner:1987vz}.  Some asymmetric $(0,2)$ Gepner 
models have been considered in the past, as heterotic compactifications with non-standard gauge 
bundles~\cite{Schellekens:1989wx,Blumenhagen:1995tt,Berglund:1995dv,Blumenhagen:1995ew,Blumenhagen:1996gz}. In contrast type IIA/IIB 
asymmetric Gepner models have not been explored in detail;  as we shall see they provide 
a good starting point for constructing large classes of non-geometric backgrounds. 

Discriminating between abstract worldsheet theories with and without a geometrical target-space interpretation is quite difficult. We shall 
use in this work a simple sufficient (but not necessary) criterion.   Let us consider a compactification of type IIA or type IIB 
superstrings (without orientifolds, D-branes or RR fluxes) such that all space-time supersymmetry comes from the left-moving worldsheet degrees of freedom.  A {\it geometric} 
compactification of this sort would exist if the two connections with torsion $\nabla \left(\omega \pm H/2\right)$ appearing in 
the supersymmetry variations of the gravitini gave  different G-structures, requiring non-zero three-form flux; for compact models this 
is forbidden by the tadpole condition quoted above.

The inspiration for this  article originates from a recent work~\cite{Israel:2013wwa} where we described fibrations of 
$K3$ Gepner models over a two-torus in type II, breaking space-time supersymmetry from the right-movers only. Following 
our general argument it implies that, while going around a one-cycle of the base, the $K3$ fiber undergoes a non-geometric symmetry 
twist. The symmetries of the  $K3$ fiber appearing in the monodromies are actually neither geometric symmetries nor mirror 
symmetry. These new non-geometric symmetries of CY quantum sigma-models are the focus of the present work. We shall embed them 
in a larger framework of non-geometric models based on solvable  Calabi-Yau compactifications.

We construct a wide class of asymmetric Gepner models in type II, using the simple currents 
formalism~\cite{Schellekens:1990xy}, preserving space-time supersymmetry from the left-movers,  
while the other half is generically broken. This is made possible by a specific choice of discrete torsion, which changes 
in particular the orbifold action on the K\"ahler moduli. This leads to many non-geometric compactifications with $\mathcal{N}=1$ 
supersymmetry in four dimensions, and a reduced moduli space of vacua. We will present some examples based on the quintic to illustrate 
these features. 

In some cases, including those underlying the $K3$ fibrations over $T^2$ studied in~\cite{Israel:2013wwa}, the non-geometric fluxes lead to superconformal 
field theories isomorphic to the original ones, albeit of a different nature. These are generalizations of 
mirror symmetry of $(2,2)$ models (in heterotic strings,  $(0,2)$ extensions of mirror symmetry have  
been considered  in~\cite{Blumenhagen:1996vu} and subsequent works),  that take the sigma-models out of the realm of CY compactifications.  

Mirror symmetry plays a major role in our understanding of CY manifolds, both in their physical and mathematical 
aspects~\cite{Yau:1998yt,Greene:1997ty}. It generalizes  T-duality to CY sigma-models 
as one exchanges the axial and vector R-symmetries of the superconformal 
algebra~\cite{Lerche:1989uy}. The first concrete realization was obtained by Greene and Plesser~\cite{Greene:1990ud} using Gepner 
models; they have shown that an orbifold by the largest subgroup of discrete symmetries preserving 
spacetime supersymmetry (bar permutations) gives an isomorphic conformal field theory with reversed right-moving R-charges. 
It provides an equivalence between type IIA compactified on some CY and type IIB  on a 
topologically distinct one, whose Hodge diamonds are 'mirror' to each other.  
Using the gauged linear sigma-model (GLSM) description~\cite{Witten:1993yc}, that includes the Gepner points, 
Hori and Vafa gave a proof of mirror symmetry~\cite{Hori:2000kt}; in this context the dual models appear naturally 
as orbifolds of Landau-Ginzburg (LG) models.

As a special case of the general construction of asymmetric Gepner models that we present in this work, there exists 
a subclass of models such that the axial and vector 
R-symmetries for a {\it single} minimal model are exchanged; they are isomorphic as CFTs to the original theory. 
They correspond to 'hybrid'  LG orbifolds with both chiral and twisted chiral superfields, hence cannot be lifted 
to ordinary Calabi-Yau GLSMs; this is a sign of the non-CY nature of these new dual models. Following the ideas 
of~\cite{Hori:2000kt} we will propose a hybrid GLSM that provides their UV completion. Given that the map can be applied 
stepwise to each and every minimal model until we reach the usual mirror theory, we give to this symmetry the name of 
{\it fractional mirror symmetry}.  

This work is organized as follows. In section~\ref{sec:gepreview} we present a short overview of simple currents and Gepner 
models orbifolds. In section~\ref{sec:nongeomCY} we provide the general construction of $\mathcal{N}=1$ non-geometric 
compactifications, and study some explicit examples based on the quintinc. In section~\ref{sec:fracmirror} 
we define and study fractional mirror symmetry. Finally in section~\ref{sec:disc} we give the conclusions and explain the 
relation between the $K3$ fibrations over tori of~\cite{Israel:2013wwa} and these new constructions. 
Useful facts about $\mathcal{N}=2$ characters and representations are given in the appendix. 

\section{Simple currents and Gepner models}
\label{sec:gepreview}
Let us first review briefly the simple current formalism~\cite{Schellekens:1989am,Kreuzer:1993tf}, its relation with Gepner 
models and orbifolds thereof. 

\subsection{Simple currents and discrete torsion}
In a conformal field theory a {\it simple current} $J$ is a primary of the chiral algebra whose fusion with a generic primary 
gives a single primary: $J \star \phi_{\mu} = \phi_{\nu}$. This action defines 
the {\it monodromy charge} of the primary w.r.t. the current, $Q_\imath (\mu)= \Delta(\phi_\mu)+\Delta({J_\imath})
-\Delta (J_\imath \star \phi_\mu)\! \mod 1$;  
two-currents are  mutually local if $Q_{\imath} (J_\jmath)=0$. We consider the extension of a rational CFT by a set of $M$ simple 
currents $J_\imath$. Provided that the simple currents action 
has no fixed points, the associated modular-invariant partition function is:
\begin{equation}
\label{eq:curpart}
Z = \sum\limits_{\mu} \prod\limits_{\imath=1}^{M} \sum\limits_{b^\imath \in \mathbb{Z}_{n_\imath}}
\chi_{\mu} (q)\, \chi_{\mu+ \beta_\jmath b^\jmath} (\bar q)  \
\delta^{(1)} \left( Q_{\imath} (\mu) + X_{\imath \jmath} b^{\jmath} \right)\, , 
\end{equation}
with $J_{\imath} \star \phi_{\mu} = \phi_{\mu+\beta_\imath}$ and $n_\imath$  the length of $J_\imath$.
The symmetric part of the  matrix $X$ is determined by the relative monodromies as 
$X_{\imath \jmath} + X_{\jmath \imath}= Q_{\imath} (J_\jmath)$, while the antisymmetric part, 
{\it discrete torsion}, should be such that:
\begin{equation}
\label{eq:torsionconst}
\text{gcd} (n_\imath,n_\jmath)\, X_{\imath  \jmath} \in \mathbb{Z}\, .
\end{equation}
If the left and right kernels of $X$ are different, the simple-current-extended modular invariant is asymmetric.

\subsection{Gepner models}
A Gepner model for type II superstrings compactified on a CY threefold is obtained from a tensor product of $r$ $\mathcal{N}=(2,2)$ minimal models,  
whose central charges satisfy \mbox{$\sum_{n=1}^r c_n = \sum_{n=1}^r (3-6/k_n)= 9$}, tensored with a 
free $\mathbb{R}^2$ superconformal theory that represents the space-time part in the light-cone gauge.  
One needs to project the theory onto states with odd integer left and right R-charges; this 
can be rephrased in the simple currents formalism.  The simple currents of the minimal models are 
primaries with quantum numbers ({\it j}=0,{\it m},{\it s}). These simple currents  
can be grouped together with the current for a free fermion into a simple current $J$ 
with labels 
\begin{equation}
\beta_J=(s_0|m_1,\ldots,m_r|s_1,\ldots,s_r)\, ,
\end{equation} 
where $s_0$ is the fermionic $\mathbb{Z}_4$ charge of the $\mathbb{R}^2$ factor. 

The Gepner modular invariant is obtained as a simple current extension, using first the 
sets of currents $\{ J_n , n=1,\ldots,r\}$, with 
\begin{equation}
\beta_{n}=(2|0,\ldots,0|0,\ldots,0,\!\!\!\!\!\!\!\underbrace{2}_{n-\text{th position}}\!\!\!\!\!\!\!,0,\ldots,0)
\end{equation}
enforcing world-sheet supersymmetry, and second the current $J_0$, with 
\begin{equation}
\beta_{0}=(1|1,\ldots,1|1,\ldots,1)\, , 
\end{equation}
ensuring the projection onto odd-integer R-charges hence  space-time supersymmetry. 
All these simple currents are  mutually local. 

In order to write the Gepner model partition function in a compact way we gather the free-fermion character $\theta_{s_0,2}/\eta$ 
and minimal models characters $\chi^{j \phantom{(s)} }_{m,\, s}$ as 
\begin{equation}
\chi^\lambda_\mu (q) = \frac{\theta_{s_0,2} (q)}{\eta (q)} \times \prod_{n=1}^r \chi^{j_n \phantom{(s_n)} }_{m_n,\, s_n} (q)\, ,
\end{equation} 
where  we have grouped the associated quantum numbers as follows 
\begin{equation}
\lambda=(j_1,\ldots,j_r) \ \text{and}  \quad \mu=(s_0|m_1,\ldots,m_r|s_1,\ldots,s_r)\, .
\end{equation}
The diagonal modular-invariant partition function of a CY$_3$ compactification at a Gepner point is then given by:
\begin{equation}
Z = \frac{1}{2^r}\frac{1}{\tau_2^2 |\eta|^4}  \sum_{\lambda, \mu} \sum_{b_0 \in \mathbb{Z}_K} 
(-1)^{b_0}\ \delta^{(1)} \left(\frac{Q_R-1}{2} \right)  \, 
\prod_{n=1}^{r}  \sum_{b^n \in \mathbb{Z}_{2}}\delta^{(1)} \left(\frac{s_0-s_n}{2} \right)
\chi^\lambda_\mu (q) \chi^\lambda_{\mu + \beta_0 b_0 + \beta_l b^l}(\bar q)\, \, , 
\label{eq:cy3part}
\end{equation}
where $Q_R$ is the left-moving worldsheet R-charge and  $K=\text{lcm} (2k_1,\ldots,2k_r)$. One can 
check that the right-moving R-charge $\bar{Q}_R$ takes also odd-integer values.

\subsection{Supersymmetric orbifolds and mirror symmetry}
Simple currents preserving world-sheet and space-time supersymmetry should be mutually local with respect to the 
Gepner model currents $\{J_0,J_1,\ldots,J_r\}$, see~\cite{Schellekens:1989wx}. Let us consider a 
generic simple-current $\mathfrak{J}$
with
\begin{equation}
\label{eq:generic}
\beta_\mathfrak{J}=(0|2\rho_1,\ldots,2\rho_r|0,\ldots,0)\ , \quad \rho_n \in \mathbb{Z}\, .
\end{equation}

Any such current  is mutually local w.r.t. the set of currents $\{\, J_n\, \}$, hence the corresponding 
extended partition function always preserves worldsheet supersymmetry. Mutual locality  with respect to the current $J_0$ (which 
ensures odd integrality of the R-charges) requires that 
\begin{equation}
\sum_{n=1}^r \frac{\rho_n}{k_n} \in \mathbb{Z}\, .
\end{equation}
If this condition is satisfied one obtains an $\mathcal{N}=2$ compactification, corresponding 
to a Calabi-Yau orbifold at a Gepner point. 
 
Extending a Gepner model with {\it all} such supersymmetry-preserving simple currents (without discrete torsion)  
gives the mirror Gepner model, which is such that the  right R-charge $\bar Q_R$ has opposite sign compared to the 
original model; it exchanges the chiral and twisted chiral rings of the theory, hence the complex structure and 
K\"ahler moduli spaces. This is the basis of 
the construction of mirror manifolds by Greene and Plesser~\cite{Greene:1990ud}. 

\section{Non-geometric CY compactifications}
\label{sec:nongeomCY}
In this section we will describe a way to obtain many non-geometric models starting from a Calabi-Yau 
compactification at a Gepner point. 

\subsection{General method}
In order to construct new non-geometric compactifications we consider extensions of the Gepner model partition function by simple currents that 
are {\it not} mutually local w.r.t. the Gepner model currents.  A generic current $\mathfrak{J}$ as in eq.~(\ref{eq:generic}) is 
actually non-local w.r.t. $J_0$,  hence space-time 
supersymmetry is completely broken (while worldsheet supersymmetry is preserved). Indeed
\begin{equation}
Q_0 (\mathfrak{J}) =  \sum_{n=1}^r \frac{\rho_n}{k_n} \mod 1\, .
\end{equation}

Now comes the key step; there is a choice of discrete torsion, consistent with eq.~(\ref{eq:torsionconst}) 
for any $\{\rho_n \in \mathbb{Z}\}$, given by 
\begin{equation}
\label{eq:discretetorsion}
X_{0\mathfrak{J}}^{\text{antisym}} = -\frac{1}{2} \sum_{n=1}^r \frac{\rho_n}{k_n}\, ,
\end{equation}
bringing down the $X$ matrix to a lower-triangular form. Its  only non-zero entries are
\begin{equation}
\label{eq:Xtriangle}
X_{\mathfrak{J}\mathfrak{J}} = \sum_{n=1}^r \frac{\rho_n^2}{k_n} \ , \quad X_{\mathfrak{J}0} = \sum_{n=1}^r \frac{\rho_n}{k_n} \, .
\end{equation}
This choice allows to bring back the  projection onto odd-integer left-moving  R-charges $Q_R$ into its original form. 

The modular-invariant partition function of the $\mathfrak{J}$-extended Gepner model with this choice of discrete torsion is 
given by
\begin{multline}
\allowdisplaybreaks
Z = \frac{1}{2^r}\frac{1}{\tau_2^2 |\eta|^4}  \sum_{\lambda, \mu} \sum_{b_0 \in \mathbb{Z}_K} 
(-1)^{b_0}\ \delta^{(1)} \left(\frac{Q_R-1}{2} \right)  \, \times \\ \times \,  \sum_{B \in \mathbb{Z}_N}
\delta^{(1)} \left( \sum_{n=1}^r \frac{\rho_n (m_n+b_0 + \rho_n B)}{k_n}\right)
\prod_{n=1}^{r}  \sum_{b^n \in \mathbb{Z}_{2}}\delta^{(1)} \left(\frac{s_0-s_n}{2} \right)
\chi^\lambda_\mu (q) \chi^\lambda_{\mu + \beta_0 b_0 + \beta_l b^l + \beta_{\mathfrak{J}} B}(\bar q)\, \, , 
\label{eq:asycy3part}
\end{multline}
where $Q_R$ is the left-moving worldsheet R-charge and the length of the simple-current is given by 
$N=\text{lcm}\, \left( \text{lcm}\,(\rho_1,k_1)/\rho_1,\ldots, \text{lcm}\, (\rho_r,k_r)/\rho_r\right)$)~\footnote{If some 
of the $\rho_n$'s vanish, the definition of $N$ has to be modified accordingly; only non-zero entries enter the formula.}.

If some levels $k_n$ are even, there may be fixed points under the simple current action, and multiplicity factors need to be added 
accordingly to the partition function. For simplicity of presentation we assume that we do not encounter this situation, 
which does not change the salient features of the construction; for instance one can take all the levels $k_n$ to be odd. 

Thanks to the discrete torsion the projection onto odd-integer worldsheet R-charges, 
given by the discrete delta-function in the first line, has been restored in the left-moving sector; 
hence space-time supersymmetry from the left-movers is preserved. This supersymmetry is generated by spectral flow 
of the left-moving $\mathcal{N}=2$ superconformal algebra as usual. 

Twisted sectors associated with the $\mathfrak{J}$-extension ($i.e.$ states with $B \neq 0$) can have  
fractional values of the right-moving worldsheet R-charge $\bar{Q}_R$. Indeed,
\begin{equation}
\label{eq:Rcharge}
\bar{Q}_R \equiv 1+2B \sum_{n=1}^r \frac{\rho_n}{k_n} \mod 2\mathbb{Z} \ ,
\end{equation}
hence space-time supersymmetry from the right-movers is generically broken; we end up with $\mathcal{N}=1$ 
four-dimensional supersymmetry.  Following our general argument given in the introduction, this construction provides a 
whole class of non-geometric quotients of CY sigma-models at Gepner points.

Naturally it is possible to consider a simple current extension by several such currents, with a discrete 
torsion of the form~(\ref{eq:discretetorsion}) for each of them; these currents may or may not be mutually 
local. Discrete torsion with no components along the 'Gepner currents' $J_0$ and $J_n$ can be added without 
breaking further spacetime supersymmetry.

\subsection{Some quintic-based examples}
We have given a method that allows to obtain type IIA or type IIB $\mathcal{N}=1$ compactifications to four dimensions, with 
neither orientifolds nor RR fluxes, starting from quite generic non-supersymmetric geometric orbifolds of Calabi-Yau compactifications at  
Gepner points (this can be extended to a wider class of Landau-Ginzburg orbifolds, 
see section~\ref{sec:disc}). The moduli spaces of such models are  significantly reduced compared to the original 
$\mathcal{N}=2$ CY compactifications. 

To illustrate this general construction, let us consider several examples based on the quintic Calabi-Yau. 
The quintic is given by 
the hypersurface 
\begin{equation}
\label{eq:defquintic}
Z_1^{\, 5}+Z_2^{\, 5}+Z_3^{\, 5}+Z_4^{\, 5}+Z_5^{\, 5}=0
\end{equation} 
in $\mathbb{P}^4$, the complex variables $Z_n$ being homogeneous coordinates on the complex projective space. This Calabi-Yau has 
a unique complexified K\"ahler modulus $t$, whose real part is the volume modulus inherited from the ambient $\mathbb{P}^4$. 
The complex structure moduli correspond to deformations of eq.~(\ref{eq:defquintic}) by monomials of degree five; there are 101 inequivalent of them. 

In the regime $\Re (t)\to -\infty$, a quantum $(2,2)$ non-linear sigma-model (NLSM) on the quintic is described by the Gepner model 
with $k_1=\cdots=k_5=5$. Complex structure deformations correspond to marginal chiral operators of R-charges $Q_R = \bar{Q}_R=1$. They 
are obtained from a tensor product of chiral operators in each minimal model, labeled by an $SU(2)$ spin $j_n$, with 
$j_n\in \{0,1/2,\ldots,k_n/2-1\}$, such that $\sum_{n=1}^5 2j_n/k_n=1$; they correspond to the monomials 
$Z_1^{\, 2j_1}  \cdots Z_5^{\, 2j_5}$. Twisted chiral states ($i.e.$ chiral w.r.t. the left-moving 
superconformal algebra and anti-chiral w.r.t. the right-moving superconformal algebra) appear in the twisted sectors of the Gepner model 
projection, $i.e.$, with $b_0\neq 0$ in the partition 
function~(\ref{eq:cy3part}). Explicitly, the complexified K\"ahler modulus has $2j_1 =\cdots= 2j_n=1$ and $b_0=8$.

We consider simple current extensions of the form discussed in section~\ref{sec:nongeomCY}, with discrete torsion leading 
generically to $\mathcal{N}=1$ space-time supersymmetry. They are characterized by the integer-valued five-dimensional vector 
\begin{equation}
\varrho=(\rho_1,\ldots,\rho_5)\, ,
\end{equation}
giving the simple current action in each minimal model. We consider below the salient features of three representative cases.

\subsubsection{$\varrho=(1,2,3,2,1)$ model}
On top of the projection onto odd integer left R-charges, states in the partition function should satisfy the constraint
\begin{equation}
\label{eq:quintic_constr}
m_1 + 2 m_2 + 3 m_3 + 2 m_4 + m_5 + 9 b_0 + 19 B \in 5 \mathbb{Z}\, ,
\end{equation}
see eq.~(\ref{eq:asycy3part}), where the label of the twisted sectors is $B=0,\ldots,4$. For each of the 101 left chiral operators 
of charge $Q_R=1$, one can solve eq.~(\ref{eq:quintic_constr}) in a given twisted sector $B$. 

Furthermore, the right $\mathbb{Z}_{k_n}$ charge of the $n$-th minimal model is shifted in the twisted 
sectors as $m_n + b_0 \to m_n + b_0 + 2\rho_n B$. For a given spin $j_n$, chiral and anti-chiral states minimize the 
conformal dimension w.r.t. $m_n$. Therefore if $B\neq 0$ a formerly massless state could only stay
massless if, in each minimal model, either $2 \rho_n B \equiv 0 \mod 2k_n$, $i.e.$ if the shift is trivial (thanks to the periodicity 
$m_n \sim m_n + 2 k_n$), or if a formerly chiral or antichiral state becomes another anti-chiral or chiral state. 

In this particular example, we found that out of the 101 original chiral states of charge $Q_R =\bar{Q}_R=1$, corresponding to 
the complex structure deformations, only 18 operators remain massless. They all belong to the untwisted sector $B=0$ in this case, 
hence are still chiral operators; this is not a generic feature of the models as we will see in the next example.  
The former unique K\"ahler modulus of the quintic threefold is lifted, acquiring a string-scale mass 
\begin{equation}
M_K = \sqrt{\frac{2}{\alpha'}}\, .
\end{equation} 

In the original Gepner model, the gravitino corresponding to the space-time supersymmetry  
associated with the right-movers was obtained from the NSR primary operator with $2j_1=\cdots=2j_5=m_1=\cdots=m_5=0$ and 
$b_0=1$. For these quantum numbers the constraint~(\ref{eq:quintic_constr}) singles out the twisted sector $B=4$. Accordingly 
the right gravitino is now massive, with 
\begin{equation}
M_{\psi_\mu}=2\sqrt{\frac{2}{\alpha'}}\, .
\end{equation}
This mass scale is of the same order as massive string states. It indicates that one cannot reliably study this construction as a 
spontaneous $\mathcal{N}=2\to \mathcal{N}=1$ SUSY breaking from an effective $\mathcal{N}=2$ supergravity perspective.

\subsubsection{$\varrho=(0,0,1,2,3)$ model}
Only the last three minimal models are affected by the simple current extension. The untwisted sector ($B=0$) 
contains the subset of marginal chiral operators, $i.e.$ complex structure deformations, 
that are not projected out. Some of these, as 
$Z_3^{\, 2} Z_4 Z_5^{\, 2}$, involve only the last three minimal models; others as $Z_1^{\, 2} Z_2^{\, 3}$ 
involve only the first two ones; finally operators as $Z_1 Z_4^2 Z_5^2$ contain both. Overall there are 20 such marginal chiral 
operators. 

In this model, the twisted sectors ($B\neq 0$) contain also marginal operators, of a peculiar nature. While they are chiral w.r.t. 
the left-moving superconformal algebra, they are neither chiral (c) nor antichiral (a) w.r.t. the right moving one. 
In terms of the right-moving chiral or antichiral nature in the individual minimal model factors, one gets 
\begin{itemize}
\item $B=1$ sector: one operator from $Z_3 Z_4^2 Z_5^2$, of (c/a,c/a,a,c,a) chirality on the right;
\item $B=2$ sector: two operators, $e.g.$ from $Z_2 Z_3^2 Z_4 Z_5$ of (c/a,c,a,a,a) right chirality;
\item $B=3$ sector:  two operators, $e.g.$ from $Z_2 Z_3 Z_4^2 Z_5$ of (c/a,c,c,a,a) right chirality;
\item $B=4$ sector: two operators, $e.g.$ from $Z_2 Z_3 Z_4 Z_5^2$ of (c/a,c,a,c,c) right chirality.
\end{itemize}
Hence all these twisted sector marginal operators are semi-chiral operators. Anticipating the discussion of the next section, 
there is no 'duality frame' w.r.t. fractional mirror symmetry such that all massless operators are either chiral or twisted chiral.

\subsubsection{$\varrho=(0,0,0,2,2)$ model}
While analyzing the amount of space-time supersymmetry, one observes that some models preserve more supersymmetry 
than one can naively think; this example is one of them. 

As written previously, in the original Gepner model the gravitino from the NSR sector is characterized by 
$2j_1=\cdots=2j_5=m_1=\cdots=m_5=0$ and $b_0=1$. In the present case, the corresponding state belongs to the twisted sector $B=2$. 
While the first three minimal models are in the right Ramond ground state with $(j,m,s)=(0,1,1)$, 
the last two minimal models have to be in the Ramond ground state of opposite R-charge, $i.e.$ with $(j,m,s)=(0,-1,-1)$, in order to 
get a massless operator. 

In ordinary constructions of space-time supersymmetric compactifications, one imposes that the 
{\it diagonal} R-current has a spectrum of odd integer charges, such that it can be exponentiated to a spin field mutually 
local with the physical states. As eq.~(\ref{eq:Rcharge}) indicates, the model that we consider do not satisfy this property 
on the right. Nevertheless there exists a different realization of space-time supersymmetry for the 
right-moving degrees of freedom. One can check that the supersymmetry operator that we have just constructed is 
mutually local with the other operators, owing both to the left-moving GSO projection $Q_R \in 2\mathbb{Z}+1$ and to the projection 
coming from the simple current extension, namely $2(m_4+m_5+2b_0)+8B \in 5\mathbb{Z}$.

\section{Fractional mirror symmetry}
\label{sec:fracmirror}

The last example of asymmetric Gepner model that we have studied in the previous section was quite intriguing, as space-time 
supersymmetry among the right-moving degrees of freedom was realized even though the right R-charge 
$\bar{Q}_R$ was not integer-valued. In some cases the similarity between the geometric compactifications and 
the non-geometric ones goes beyond the amount of preserved supersymmetry. 

\subsection{Elementary simple current extensions and fractional mirror symmetry}

A particular type of  $\mathfrak{J}$-extensions with discrete torsion has indeed remarkable properties, as 
not only the space-time supersymmetry is the same as the original Calabi-Yau 
compactification at a Gepner point, but the whole superconformal field theories are isomorphic to each other. Extending 
a Gepner model partition function with an 'elementary' simple current of labels 
\begin{equation}
\label{eq:mirrocur}
\beta_{\mathfrak{J}}=(0|2,0,\ldots,0|0,\ldots,0)\, ,
\end{equation}
amounts, while taking into account the twisted sectors and discrete torsion, to replace in the original partition function 
the anti-holomor\-phic character for the first minimal model with the character of opposite $\mathbb{Z}_{k_1}$ charge, namely
\begin{equation}
\label{eq:map}
\chi^{j_1}_{m_1 + b_0,\, \, s_1+b_0+2b_1 }(\bar q) \ \xrightarrow{\mathfrak{J}\text{\tiny -ext.}} 
\ \chi^{j_1}_{-m_1  - b_0,\, s_1+b_0+2b_1}(\bar q)\, .
\end{equation} 
In the right NS sector, there is an equivalence between the map~(\ref{eq:map}) and changing the sign of the right R-charge associated 
with the first minimal model.  In the right Ramond sector it is also true if one changes the right-moving space-time chirality 
at the same time.  As superconformal field theories the original model and the new one are therefore isomorphic.

Starting from a type IIA Calabi-Yau compactification at a Gepner point, we obtain a type IIB theory on a 
Gepner model whose right-moving R-charge associated with the first minimal model has been reversed; with respect to the 
original right-moving diagonal R-current the spectrum of R-charges is not integer-valued hence the model 
is not associated with a Calabi-Yau. Put it differently the quotient does not preserve the holomorphic three-form. 
These two models are isomorphic  hence describe the same physics. This {\it fractional mirror symmetry} can be applied stepwise 
until one obtains the mirror description in the usual sense.

\subsection{Hybrid Landau-Ginzburg models}
A minimal model is the IR fixed point of a LG model 
with superpotential $W=Z^{k}$~\cite{Witten:1993jg}. Its mirror, obtained by a $\mathbb{Z}_k$ quotient, 
is a LG model for a twisted chiral superfield $\tilde{Z}$ with a twisted superpotential 
$\tilde{W}=\tilde{Z}^k$. In the present context we are considering a similar quotient acting inside a LG 
orbifold, with a discrete torsion that  disentangles partly the two orbifolds --~the diagonal one ensuring R-charge 
integrality and the $\mathbb{Z}_{k_1}$ quotient giving the fractional mirror. We end up with a 'hybrid' Landau--Ginzburg orbifold containing 
both a twisted chiral superfield $\tilde{Z}_1$ and chiral superfields $Z_{2,\ldots,r}$; hence it cannot be related 
to a Calabi-Yau GLSM. 

This quantum equivalence needs not be restricted to the Gepner points in the Calabi-Yau moduli space. To illustrate this point 
let us consider again the quintic. Away from the Gepner point, we expect that for every hypersurface in $\mathbb{P}^4$ of the form 
\begin{equation}
z_1^5 + \sum \alpha_{abc} z_2^{a} z_3^{b} z_4^{c} z_5^{5-a-b -c}=0
\end{equation}
a realization of fractional mirror symmetry w.r.t. the chiral superfield $Z_1$ as a simple current extension similar 
to~(\ref{eq:mirrocur}) exists. In other words the complex structure deformations that preserve the 
$\mathbb{Z}_{5}$ symmetry $z_1 \to e^{2i\pi/5} z_1$ are compatible with this construction. K\"ahler 
deformations are not compatible with this $\mathbb{Z}_{5}$ symmetry, as can be seen explicitly at the Gepner point; hence this realization of fractional mirror symmetry as 
a quotient is not present in the large-volume limit.\footnote{This is somehow similar to the Greene-Plesser realization of usual 
mirror symmetry, which involves the extra discrete symmetries of hypersurfaces of the Fermat type.} 

When these conditions are met the $\mathcal{N}=2$ superconformal algebra can be split into the algebra coming from the LG model 
$W=Z_1^{\, 5}$ and from the LG model for the other multiplets. This allows to dualize $Z_1$ into a twisted chiral multiplet, 
giving a more general 'hybrid' LG orbifold with superpotential $W=\sum \alpha_{abc} Z_2^{a} Z_3^{b} Z_4^{c} Z_5^{5-a-b -c}$ 
and twisted superpotential $\tilde W = \tilde{Z}_1^5$. One expects also that other accidental splittings of the superconformal 
algebra, corresponding to orbifolds of tensor products of Landau-Ginzburg models of more generic form 
(for instance if the LG potential splits as $W=G(Z_1,Z_2) + H(Z_3,Z_4,Z_5)$),  
should give rise to different fractional mirror symmetries. 

The dualities between the Calabi-Yau compactifications and their fractional mirrors presumably hold also outside of the 
Landau-Ginzburg regime, at least in some neighborhood. Indeed the duals of the K\"ahler moduli 
are {\it semi-chiral} operators, whose left conformal dimension $h$ are protected. Given that the spin 
$h-\bar h$ should be integer, their right conformal dimension $\bar h$ cannot vary continuously, hence is 
fixed assuming that no jumps occur.

\subsection{Fractional mirror symmetry and GLSMs}

On the Calabi-Yau side, Landau-Ginzburg orbifolds and Calabi-Yau NLSMs  describe the infrared dynamics of $(2,2)$ gauged 
linear sigma-models in different regimes, continuously connected by varying the Fayet-Iliopoulos parameters, $i.e.$  
giving vacuum expectation values to marginal twisted chiral operators in the infrared description.  
It would be very helpful to have then a 'UV completion' of the dual theory, which can 
be taken out of the 'hybrid' Landau-Ginzburg regime where both formulations become overtly equivalent. We shall propose 
below such description.

For concreteness we consider again the $(2,2)$ gauged linear sigma-model for the quintic three-fold. This two-dimensional gauge theory 
contains a $U(1)$ vector superfield, five chiral superfields $Z_{1,\ldots,5}$ of charge one and a chiral superfield $P$ 
of charge $-5$. They interact through the superpotential $W=P\, G(Z_n)$, where $G=\sum Z_{n}^{\, 5}$ is the degree five homogeneous 
polynomial defining the CY hypersurface; furthermore the theory contains a twisted superpotential $\tilde{W}=-t \, \Sigma$, $t=r-i\theta$ 
being the complexified Fayet-Iliopoulos parameter and $\Sigma$ the field-strength superfield. In the regime $\Re (t) \ll 0$,  
or equivalently when $p  \neq 0$, it flows to the diagonal $\mathbb{Z}_5$ orbifold of the Landau-Ginzburg model with 
superpotential $W=G(Z_n)$. 

Using the approach of Hori and Vafa to mirror symmetry~\cite{Hori:2000kt}, one can dualize only 
the chiral superfield $Z_1$ to a twisted chiral superfield $\tilde{Y}$. 
One expects to get a 'hybrid' GLSM with superpotential and twisted superpotential 
(with $\hat{G}=Z_2^{\, 5} + \cdots Z_5^{\, 5}$):
\begin{equation}
\label{eq:superpot}
W = P \, \hat{G}(Z_n) \ , \quad \tilde{W}= \Sigma (\tilde{Y}-t) +  e^{-\tilde{Y}}\, ,
\end{equation}
where the second term in the twisted superpotential comes from worldsheet instantons.

Following for instance the discussion in~\cite{Adams:2006kb}, a geometrical NLSM regime of this model, if 
it existed, would be characterized by non-zero three-form flux $H$; this seems at odds with the tadpole 
condition recalled in the introduction. To settle this potential issue let us analyze classically what are 
the predictions of the GLSM corresponding to~(\ref{eq:superpot}). The vacua are determined by the scalar potential:
\begin{multline}
\label{eq:scalpot}
V(z_n,\tilde{y},p,\sigma) = \left|\hat{G}(z_n)\right|^2 + |p|^2 \sum_{n=2}^5 \left|\partial_n \hat{G}(z_n)\right|^2  + 
\left|\sigma - e^{-\tilde{y}} \right|^2\\
+\frac{e^2}{2} \left(|z_2|^2+ \cdots + |z_5|^2 - 5 |p|^2+ \Re (\tilde{y}) - \Re (t) \right)^2
+ |\sigma|^2 \left(|z_2|^2 + \cdots + |z_5|^2 + 5^2 |p|^2\right)\, .
\end{multline}
A geometrical 'phase' would be characterized by $|z_i| \neq 0$. It implies that $p=0$ by transversality of $\hat G$, 
and also that $\sigma=0$ in the vacuum as the superfields $Z_n$ are minimally coupled to the vector superfield, see the last term 
in~(\ref{eq:scalpot}). The twisted F-term (third term) shows a runaway behavior as the vacuum is 
obtained for $\Re (\tilde{y}) \to + \infty$. As a consequence the D-term condition (last but one term) cannot 
be satisfied. Hence this two-dimensional theory has no regime with a NLSM description; this result, which is 
in accordance with the supergravity tadpole condition, relies crucially on the worldsheet instanton contribution 
in $e^{-\tilde{Y}}$ to the twisted superpotential.

On the contrary there exists a hybrid Landau-Ginzburg 'phase'. Setting $p\neq 0$, which 
breaks spontaneously the gauge group to $\mathbb{Z}_5$, implies that $z_n=0$ by 
transversality of $\hat{G}$ and that $\sigma=0$. Then the twisted F-term enforces $\Re (\tilde{y}) \to + \infty$ as above. Finally, the D-term 
condition shows that $|p|$ is driven to very large values. The effective superpotential for the chiral superfields $Z_n$ is then 
of the form $W\sim \hat{G}(Z_n)$. Regarding the twisted chiral superfield, it was argued in~\cite{Hori:2000kt} that the fundamental field 
after the duality in such a compact model is given by $e^{-\tilde{Y}}=\tilde{X}^5$. This conjecture was tested in the context 
of 'ordinary' mirror symmetry by computing the BPS masses of A- or B-type boundary states; the same argument 
cannot be used in  the hybrid models, but we expect that the same field redefinition can be  carried over. 
It is not single valued, being invariant under $\tilde{X} \to e^{2i\pi/5} \tilde{X}$, 
hinting towards the orbifold structure that we obtained in the hybrid Landau-Ginzburg description. Clearly, a better understanding of 
the low-energy dynamics of this theory would be helpful in making the correspondence between the hybrid GLSMs and the hybrid LG models 
more precise.

\section{Conclusions and discussion}
\label{sec:disc}

In this work we have constructed a wide class of compactifications of type IIA and type IIB superstring theories,  
starting from Calabi-Yau compactifications at Gepner points, whose generic features are $\mathcal{N}=1$ space-time 
supersymmetry (with neither orientifolds nor fluxes) and a reduced moduli space of vacua. 

As was explained in the introduction, such compactifications preserving only $\mathcal{N}=1$ four-dimensional supersymmetry are 
necessarily non-geometric, as the ten-dimensional supercharges, related respectively to the left-moving and right-moving 
worldsheet degrees of freedom, are not on the same footing. Technically, the origin of this non-geometrical nature was the introduction 
of a very specific discrete torsion, whose role was to turn a non-supersymmetric Gepner model orbifold into an $\mathcal{N}=1$ theory. 

One may argue that, after all, these models are 'almost' geometric as the discrete torsion only plays a role in the twisted sectors. 
This is not actually correct, as the tensor product of minimal models becomes a CY sigma-model at a Gepner point 
only after the extension by the 'Gepner currents'  $J_0$ and $J_n$ has been implemented. 
The discrete torsion has an effect in the twisted sectors of the $J_0$-extension, giving the compactification a 
non-geometric nature. In particular the quotient has a different action on twisted chirals, $i.e.$ on K\"ahler moduli, compared to 
the corresponding geometric orbifold.

As a special case of this construction, we have obtained new quantum symmetries associated with 
superconformal field theories lying  the moduli space of Calabi-Yau compactifications, that we have called 
{\it fractional mirror symmetry}. Unlike the usual mirror symmetry which is understood everywhere in the CY moduli space, 
these new dualities are visible only when accidental discrete symmetries become manifest, in the Landau-Ginzburg regime. We have 
proposed a gauged linear sigma-model description of the dual theory, that provides a UV completion and 
can be taken out of this regime but does not exhibit a geometrical 'phase', as expected.

The asymmetric $K3$ fibrations over $T^2$ that we have given in~\cite{Israel:2013wwa} can be rephrased 
in light of the construction exposed in this article. These models, that we obtained considering some  
modular properties of $\mathcal{N}=2$ characters,  can be interpreted as fibrations  
of $K3$ at Gepner points over a two-torus, with a non-geometric monodromy twist around each 
one-cycle of the base. For this purpose one considers two 'elementary' $\mathfrak{J}$-extensions as~(\ref{eq:mirrocur}), 
acting respectively in the first and second factors  of a $K3$ Gepner model,  and as $\mathbb{Z}_{k_{1}}$ and 
$\mathbb{Z}_{k_{2}}$ shifts along the two-torus. These models are close relatives of 
T-folds~\cite{Dabholkar:2002sy} and interpolate between the $K3$ sigma-model in the large torus limit, 
and a 'half-mirror' K3 in the opposite small volume limit. As the worldsheet realization of space-time supersymmetry 
on the right-moving side is different in the theories appearing in these two limits, 
the interpolating model naturally breaks this half of space-time supersymmetry.  Furthermore, as shown  
in~\cite{Dabholkar:2002sy}, at the minimum of the four-dimensional supergravity potential, which is where 
the on-shell worldsheet description is defined, only fields invariant under the symmetry used in the twisting stay massless. 
It explains that, in many cases, all the K3 moduli are lifted in this construction. As there are no massless Ramond-Ramond fields 
in these models, it is possible to add D-branes alone to these compactifications without running into a problematic RR tadpole. 
One expects that the associated open string spectra are non-supersymmetric, yet the potential 
phenomenological implications of such models are worth exploring in detail. 

Finally it would be interesting to find whether these symmetries are related to the {\it Mathieu moonshine}, which  
suggests that $K3$ compactifications have an underlying $M_{24}$ symmetry whose origin is not fully 
understood~\cite{Eguchi:2010ej}.

\subsection*{Note added after publication}
As was realized by the author after publication, a discrete torsion similar to that 
discussed in Sec. IIIA of the paper, whose role is to restore space-time supersymmetry otherwise broken by a nonsupersymmetric 
quotient of a (2,2) Gepner model, was described already by Intriligator and
Vafa in the context of Landau-Ginzburg orbifolds in~\cite{Intriligator:1990ua} (related observations, for (0,2) models, 
were also made in~\cite{Font:1989rs}). In the present work, a similar mechanism is described, using the formalism of 
simple currents of rational conformal field theories, leading to nongeometric type II compactifications.
The main goals of the paper, besides discussing these asymmetric orbifolds of Calabi-Yau compactifications, are the
following:
\begin{enumerate}
\item to show accidental enhancements of space-time supersymmetry, with a nonstandard realization on the world sheet
which is suitable for those theories with noninteger right R charges in their spectra (Sec. IIIB)
\item to define fractional mirror symmetries (that arise from this construction as specific classes of models) and constitute
an interesting new type of symmetry between string compactifications, relating geometric and nongeometric
ones (Sec. IV)
\item to provide a realization of the models dual to the Calabi-Yau ones through fractional mirror symmetry in terms of a
new type of hybrid gauged linear sigma model, with no geometrical phase (Sec. IVC)
\item to interpret the asymmetric Gepner models that we constructed in~\cite{Israel:2013wwa}, 
which lift all K3 moduli, as nongeometric K3 fibrations over $T^2$ bases (Sec. V).
\end{enumerate}
All of this, as well as the study of massless spectra, constitutes new and original work.

\begin{acknowledgments}
I thank Ilka Brunner, Atish Dabholkar, Stefan Groot Nibbelink, Nick Halmagyi, Jan Louis, Michela Petrini and Jan Troost for 
discussions. This work was conducted within the ILP LABEX (ANR-10-LABX-63) supported by French state funds managed by the
ANR (ANR-11-IDEX-0004-02) and by the project QHNS in the program ANR Blanc SIMI5 of Agence National de la
Recherche. 
\end{acknowledgments}

\appendix*
\section{$\boldsymbol{{\mathcal N}=2}$ characters}
\label{appchar}
The characters of the $\mathcal{N} =(2,2)$ minimal model with $c=\bar c=3-6/k$, {\it i.e.}  
the supersymmetric $SU(2)_k / U(1)$ gauged \textsc{wzw} model, 
are conveniently defined through the characters $\chi^{j \phantom{(s)} }_{m,\, s}$ of the $[SU(2)_{k-2} \times U(1)_2] / U(1)_k$ bosonic 
coset, obtained by splitting the Ramond and Neveu-Schwarz 
sectors according to the fermion number mod 2. Defining $q=e^{2\pi i\tau}$ and $z=e^{2\pi i\nu}$, 
these characters are determined implicitly through the
identity:
\begin{equation} \label{theta-su2}
\chi_{k-2}^{j} (\nu|\tau)
\theta_{s,2}(\nu-\nu'|\tau) = \sum_{m \in \zi_{2k}}  \chi^{j \phantom{(s)} }_{m,\, s} (\nu'|\tau)  
\theta_{m,k} \big(\nu-\tfrac{2\nu'}{k}\big|\tau\big) \, ,
\end{equation}
in terms of the theta functions of $\widehat{\mathfrak{su}(2)}_k$:
\begin{equation}\label{thSU2}
\theta_{m,k} (\tau,\nu) = \sum_{n}
q^{k\left(n+\tfrac{m}{2k}\right)^2}
z^{k \left(n+\tfrac{m}{2k}\right)}\,,  \qquad m \in \mathbb{Z}_{2k} 
\end{equation}
and $\chi^j_{k-2} $ the characters of the affine algebra $\widehat{\mathfrak{su}(2)}_{k-2}$:
\begin{equation}\label{su2-char}
\chi_{k-2}^j (\nu|\tau) = \frac{\theta_{2j+1,k} (\nu|\tau)-\theta_{-(2j+1),k} (\nu|\tau)}{i\vartheta_1(\nu|\tau)}\,.
\end{equation}
Highest-weight representations are labeled by  $(j,m,s)$, corresponding to primaries of 
$SU(2)_{k-2}\times U(1)_k \times U(1)_2$. The following identifications apply:
\begin{equation}
\label{symMM}
(j,m,s) \sim (j,m+2k,s)\sim
 (j,m,s+4)\sim
 \big(\tfrac{k}{2}-j-1,m+k,s+2\big)
\end{equation}
as  the selection rule $2j+m+s =  0  \mod 2$. The half-integer modded spin $j$ is restricted to $0\leqslant j \leqslant \tfrac{k}{2}-1$.  
The conformal weights of the superconformal primary states are:
\begin{subequations}
\label{dimchiralMM}
\begin{align}
\Delta &=  \frac{j(j+1)}{k} - \frac{m^2}{4k} + \frac{s^2}{8}\ & \text{for} & \ -2j \leqslant m-s \leqslant 2j \\
\Delta &=  \frac{j(j+1)}{k} - \frac{m^2}{4k} + \frac{s^2}{8} + \frac{m-s-2j}{2}
\ & \text{for} & \  2j \leqslant m-s \leqslant 2k-2j-4
\end{align}
\end{subequations}
and their $R$-charge reads:
\begin{equation}
Q_R = -\frac{s}{2}+\frac{m}{k} \mod 2 \,. 
\end{equation}
\begin{itemize}
\item \emph{Chiral primary states} are obtained for $m=2j$ and 
$s=0$ (thus even fermion number). Their conformal dimension reads:
\begin{equation}
\Delta= \frac{Q_R}{2} = \frac{j}{k}\, .
\end{equation}
Equivalently they are of the form $(j,m,s)=(j,-2(j+1),2)$. 
\item \emph {Anti-chiral primary states} are obtained for $m=2(j+1)$ and $s=2$ 
(thus odd fermion number). Their conformal dimension reads:
\begin{equation}
\Delta= -\frac{Q_R}{2} = \frac{1}{2}-\frac{j+1}{k}\, .
\end{equation}
Equivalently they are of the form $(j,m,s)=(j,-2j,0)$. 
\end{itemize}

\bibliography{bibmirror2}

\end{document}